\documentclass[letterpaper, 10 pt, conference]{ieeeconf} 

\IEEEoverridecommandlockouts                                                   
\usepackage{mathtools, cuted}
\usepackage{graphicx}
\usepackage{amsmath,amssymb}
\usepackage{subfigure} 
\usepackage{array}
\usepackage{lipsum}
\usepackage{mathtools}
\usepackage{booktabs}
\usepackage{mwe}
\usepackage{xcolor}
\usepackage{float}
\makeatletter
\def\endthebibliography{%
	\def\@noitemerr{\@latex@warning{Empty `thebibliography' environment}}%
	\endlist
}
\makeatother
\newtheorem{defi}{Definition}
\newtheorem{assume}{Assumption}
\newtheorem{remm}{Remark}
\newtheorem{prop}{Proposition}

\newtheorem{thm}{Theorem}[section]

\title{\LARGE \bf
Dendritic trafficking: synaptic scaling and structural plasticity
}
\author{Saeed Aljaberi $^{1}$,  Timothy O'Leary $^{1}$, Fulvio Forni$^{1}$ 
\thanks{*S. Aljaberi is supported by Abu Dhabi National Oil Company (ADNOC). T. O'Leary is supported by ERC grant StG 716643 FLEXNEURO.}
\thanks{$^{1}$The authors are with the Department of Engineering, University of Cambridge, Cambridge CB2 1PZ, U.K.
        {\tt\small sa798|timothy.oleary|f.forni@eng.cam.ac.uk}}
}

\begin{document}

\maketitle
\thispagestyle{empty}
\pagestyle{empty}

\begin{abstract}
Neuronal circuits internally regulate electrical signaling via a host of homeostatic mechanisms. Two prominent mechanisms, synaptic scaling and structural plasticity, are believed to maintain average activity within an operating range by modifying the strength and spatial extent of network connectivity using negative feedback. However, both mechanisms operate on relatively slow timescales and face fundamental limits due to phase lags. We show that these mechanisms fulfill complementary roles in maintaining stability in a large network. In particular, even relatively, slow growth dynamics improves performance significantly beyond synaptic scaling alone.

\end{abstract}

\section{INTRODUCTION}
Neurons are excitable cells that can interconnect with thousands of other cells. The excitable properties of neurons as well as the synaptic connections between them depend on ion-permeable channels and receptor proteins that have relatively short half-lives and thus require continual replenishment \cite{marder2006variability}. Furthermore, learning and memory in biological neuronal networks is implemented by adaptation of the densities of these signaling molecules at synapses as well as formation and elimination of synaptic connections. Such adaptation in connectivity can potentially lead to destabilization of activity in large networks. Homeostatic synaptic scaling, a form of synaptic plasticity, normalizes inputs to keep neuronal activity within an operating range \cite{turrigiano2008self}. 

However, the complex geometry of neurons impose significant constraints on synaptic scaling and ion channel homeostasis. Channel mRNAs and protein subunits need to be actively transported via protein motors on cytoskeletal components called microtubules \cite{nirschl2017impact,kapitein2011way,bressloff2009cable}. The synthesis process of ion channels and receptor proteins depends on a feedback signals, including calcium influx \cite{o2010homeostasis,o2014cell}. Previous work showed that transport processes suffer from severe delays due to neurite length \cite{williams2016dendritic}, and can pose a potential source of instability in the presence of cellular feedback control \cite{aljaberi2019qualitative,aljaberi2021dendritic}.

During development, neural activity strongly regulates the growth of neurites themselves. Unlike synaptic scaling, such structural plasticity encompasses morphological changes, such as outgrowth/shrinking of dendrites or axons, dendritic branching, and formation/elimination of synapses \cite{butz2009activity}. Such changes in the neuron's structural properties happen in response to variations in the electrical activity \cite{mcallister2000cellular}. Thus, there is strong coupling between average activity and neurite growth. These changes are particularly prevalent during development as neurites grow and as connections first form. Thus, alterations to this process can profoundly and permanently affect the function of a mature network \cite{giachello2017regulation,turrigiano2004homeostatic}.

Both synaptic scaling and structural plasticity are crucial homeostatic processes that remain poorly understood at a system theoretic level. During development there is the potential for pathological instabilities to arise in neural activity, so it is important to understand how these homeostatic mechanisms interact and whether they can produce pathological states. We address this by formulating a system theoretic model that captures both processes from first principles. We derive conditions under which these mechanisms guarantee stability. We derive a necessary condition for homeostasis, that is, stability of a network, that relates biosynthesis rates of signaling proteins to morphological parameters of neurons (typical neurite length). Finally, we show that structural plasticity provides a stabilizing influence when synaptic scaling is aggressive. 

In Section \ref{sec:syn_scaling} and \ref{sec:growth_adapt} we develop a model of synaptic scaling as well as a model of activity dependent neurite growth. The interaction between synaptic scaling and activity dependent growth leads to interesting dynamical behavior including potential instabilities. Stability and general properties of this system are analyzed in Section \ref{sec:cl}.
Conclusions and future research directions follow. The proof of the main theorem is in appendix.

\begin{figure}[htpb]
	\centering
	\includegraphics[width=1.0\columnwidth]{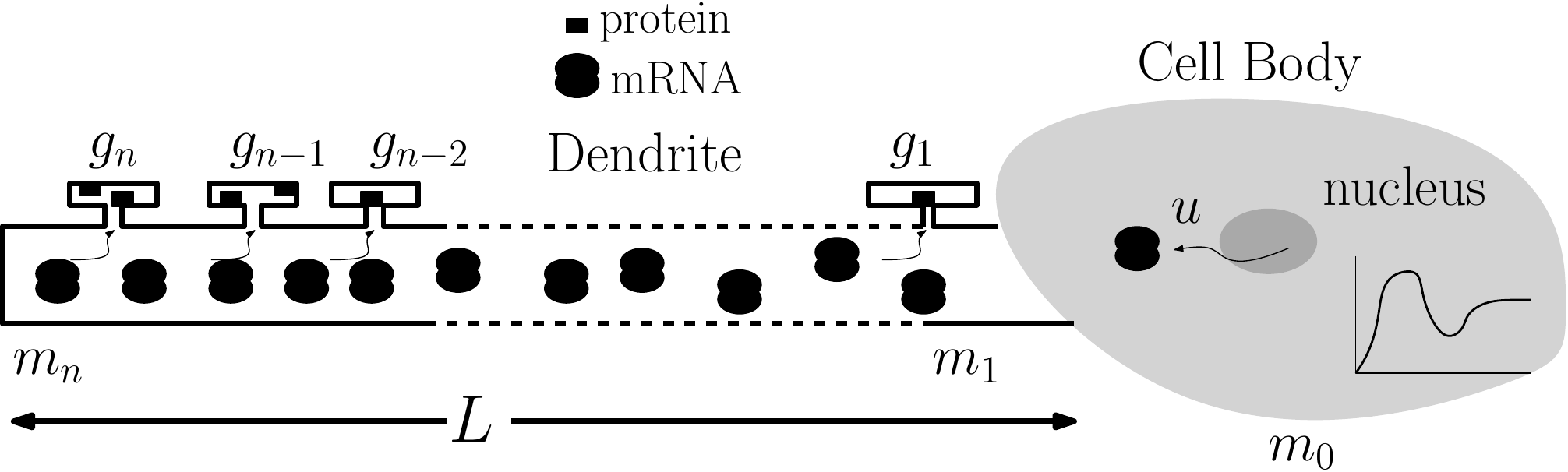}
	\caption{Cargo trafficking and ion-protein synthesis. }
	\label{fig:neuron}
\end{figure}

\section{Synaptic scaling}
\label{sec:syn_scaling}
Ion channels and receptors are responsible for the electrical activity of the neuron and are distributed throughout the cell. These signaling components are continually replenished through synthesis and transport as intracellular cargo. A common kind of cargo is mRNA that is used to synthesize ion channel proteins locally in dendrites.
The concentration of these cargo is regulated by feedback mechanisms, typically based on the 
average neural activity which correlates with calcium concentration.
In what follows we provide a basic model of these three main components, namely 
cargo trafficking, ion-protein synthesis, and feedback regulation\footnote{It is also possible that the ion-channel 
 proteins are synthesized directly in the cell-body then transported to the synapses, where they are inserted.
 This would not introduce fundamental modifications to our model.}.
 
\subsection{Cargo trafficking}
We model cargo trafficking as a \emph{nonlinear active transport phenomenon}. 
Unlike diffusion, which depends on concentration gradients, active transport requires energy and 
is performed by protein motors, such as dynein, kinesin and myosin. These motors interact
with microtubules and actin filaments, which take the role of track rails. Crowding and finite size effects
must also be taken into account, typically through classical statistical physics modeling \cite{spohn2012large}
and  mean-field approximation, leading to simple nonlinear ODEs modeling saturation effects.
In this paper the complex cargo dynamics is approximated by the following compartmental
model:
 \begin{align} 
\label{EQ:m}
\dot m_{0}  & = u - m_{0}(c - m_{1}) -\omega_{m}m_{0} \nonumber \\
\dot m_{1} & =    m_{0}(c - m_{1}) +\frac{v_{b}}{c^{2}}(c - m_{1})m_{2}   - \frac{v_{f}}{c^{2}}(c - m_{2})m_{1} \nonumber \\
& -\omega_{m}m_{1} \nonumber \\
\dot m_{i} & =   - \frac{v_{f}}{c^{2}}(c - m_{i+1})m_{i} +\frac{v_{b}}{c^{2}}(c - m_{i})m_{i+1} \nonumber \\
& + \frac{v_{f}}{c^{2}}(c - m_{i})m_{i-1} 
  - \frac{v_{b}}{c^{2}}(c - m_{i-1})m_{i} - \omega_{m} m_{i} \nonumber \\
\dot m_{n}  & =  \frac{v_{f}}{c^{2}}(c - m_{n})m_{n-1} - \frac{v_{b}}{c^{2}}(c - m_{n-1})m_{n} - \omega_{m} m_{n}.
\end{align}
\eqref{EQ:m} is a basic nonlinear transport dynamics. From Figure \ref{fig:neuron},
$m_{0}$ represents the mRNA concentration in the soma, $m_{i} \in [0,c]$ represents mRNA concentration in dendritic compartment $i \ge 1$, $v_{f}$ and $v_{b}$ are the forward and backward transport rates, respectively, $\omega_{m}$ is the mRNA degradation rate, $u$ represents mRNA production, and $c$ represents
the finite capacity of a single compartment, to model crowding. 
We assume crowding effects do not occur in the soma, since it is significantly larger than dendritic compartments. 

\begin{figure}[htpb]
	\centering
	\vspace*{3mm}
	\includegraphics[width=1.0\columnwidth]{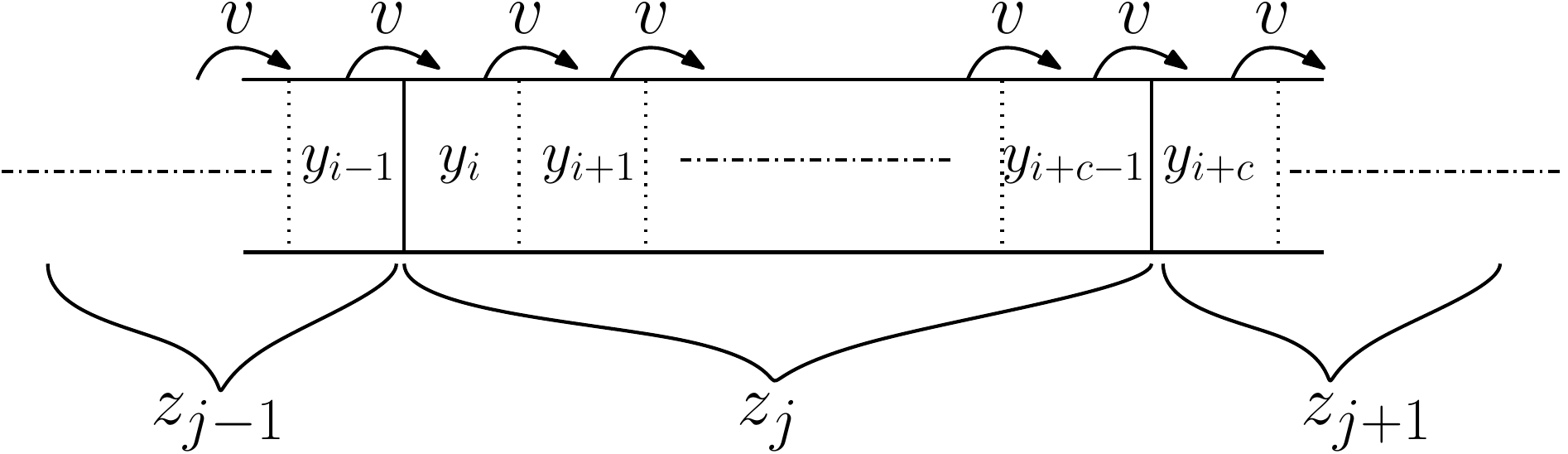}
	\caption{Microscopic picture of transport.}
	\label{fig:sch}
\end{figure}

Notably, the transport rates $v_{f}(c - m_{i})$ and $v_{b}(c - m_{i})$ are scaled by the square of the capacity, $c^{2}$.
The purpose is to adapt transport rates in accordance with the compartment size. In contrast to increasing the number
of compartments, this enables modeling of growth by adaptation of capacity and transport rates.
For instance, consider forward transport with normalized capacity $c=1$. 
A large compartment with cargo denoted by 
 $z_{j}$ can be modeled as a collection of smaller 
 compartments whose cargo is denoted by $y_{i}$, as shown in Figure \ref{fig:sch}. 
 From \eqref{EQ:m} we get the generic transport dynamics
\begin{align*}
\dot z_{j} & =  vy_{j-1}(1 - y_{i}) - vy_{i+c-1}(1 - y_{i+c}), 
\end{align*}
where the internal exchange of molecules sums to zero. Focusing on the molecules that enter and leave $z_{j}$, 
and assuming that the molecules are homogeneously distributed throughout the compartment $z_{j}$ (well-mixed), we
can write 
\begin{align*}
y_{i}=y_{i+1}=\dots=y_{i+c-1}=\frac{z_{j}}{c} \ .
\end{align*}
Thus, substituting the latter  in the equation of $\dot z_{j}$, we get
\begin{align}
\label{EQ:z-scaled}
\dot z_{j} & =  v\frac{z_{j-1}}{c}\left(1 - \frac{z_{j}}{c}\right) - v\frac{z_{j}}{c}\left(1 - \frac{z_{j+1}}{c}\right) \nonumber \\
& =  \frac{v}{c^{2}}z_{j-1}\left(c - z_{j}\right) - \frac{v}{c^{2}}z_{j}\left(c - z_{j+1}\right). \nonumber
\end{align}
which justifies the model in \eqref{EQ:m}.
\subsection{Ion channel synthesis and electrical activity}

The synthesis of ion channels $g_i$ 
from spatially distributed mRNA concentrations $m_i$ 
is a complex biochemical process that we approximate as a first order process
\begin{equation}
\label{EQ:g} 
\dot g_{i}  =  s_{i}m_{i} - \omega_{g}g_{i} \ ,
\end{equation}
where $s_{i}$ is the translation rate in compartment $i$, and $\omega_{g}$ is the degradation of ion channel proteins.
The aim of \eqref{EQ:g} is to capture roughly the temporal features and the static gain of this complex process. 

Given the timescale separation between electrical activity and transport dynamics, the effect of ion-channels $g_{i}$ on a neuron's electrical activity
can be modeled by a simple leaky-integrator model, based on a single-compartment, Ohmic current balance relationship:
$C\dot{V} = g_{leak}(E_{leak} - V) + g_{avg}(E_{g} - V)$. $V$ is the membrane potential, $C$ is the membrane capacitance, $g_{leak}$ is a fixed leak conductance, $E_{leak}$, $E_{g}$ are equilibrium potentials, and  
\begin{align}
\label{EQ:g_avg}
	g_{avg} = \frac{1}{n} \sum\limits_{i=1}^{n} g_{i}
\end{align}
is the averaged sum of ion-channel conductances. Since voltage fluctuations occur on a timescale that is
faster than cargo trafficking $m_{i}$ and protein synthesis $g_{i}$, 
we resolve the membrane voltage $V$ to a quasi-steady state 
\begin{equation}
\label{EQ:V}
V = \frac{ g_{avg}E_{g} + g_{leak}E_{leak}}{g_{leak} +  g_{avg}}. 
\end{equation}
Finally, following \cite{o2013correlations}, we observe that the electrical activity of the neuron
affects calcium concentration, captured by a sigmoidal monotone relation. For simplicity, we take 
\begin{equation*}
\label{EQ:ca}
[Ca^{+2}]  = \frac{\alpha}{1 + \exp(-V/\beta)} 
\end{equation*}
where $\alpha$ and $\beta$ shape the sensitivity of $[Ca^{+2}]$ to $V$. 
In what follows we will use the non-increasing function $h$ 
\begin{equation}
\label{EQ:ca2}
[Ca^{+2}]  = h(g_{avg}) 
\end{equation}
to denote the composition of  voltage $V$ and calcium $[Ca^{+2}]$ equations,
as illustrated in Figure \ref{fig:stat_map}.

\subsection{Ion channel / cargo regulation}

Homeostatic control of neural activity is achieved by regulation of intracellular calcium $[Ca^{+2}]$ to a certain target $ [Ca^{+2}]_{target}$.
This is achieved by controlling the production of mRNA in the soma in agreement to a (leaky) integral control law
\begin{align}
  \label{EQ:u}
  \dot{u} &= k_{u}e - \omega_{u}u
 \end{align} 
 where $e = [Ca^{+2}]_{target} - [Ca^{+2}]$, $k_{u}$ is the integral gain, 
 and $\omega_{u}$ is the degradation rate, typically small. As for the protein 
 synthesis in the previous section, feedback control is achieved through a complex biochemical process.
 A detailed model of this process is beyond the scope of this paper. The interested reader is referred to \cite{o2013correlations} and references therein.
 \eqref{EQ:u} provides a basic approximation describing timescale and steady-state gain of this complex process.
 
 Feasibility of the steady state $[Ca^{+2}] = [Ca^{+2}]_{\text{target}}$
 for the closed loop \eqref{EQ:m}-\eqref{EQ:u} is guaranteed by the following assumption.
\begin{assume}
\label{assume:feasibility}
	The parameters of \eqref{EQ:m}-\eqref{EQ:u} satisfy 
	$$
	E_{g} + \beta\ln \frac{\alpha}{[Ca^{+2}]_{\text{target}}} \ne 0 \mbox{ and } g_{leak} \ne 0 \ . \vspace{2mm}
	$$
	
\end{assume}
The closed loop given by Equations \eqref{EQ:m}-\eqref{EQ:u} describes the so-called phenomenon of synaptic scaling, a global homeostatic mechanism that prevents activity from building up and eventually reaching  pathological levels. 
The neuron, via feedback, reduces/increases the cargo density in the system, which in turn reduces/increases the ion-channel density, ultimately normalizing the electrical activity of the neuron. For a detailed study of this feedback mechanism
the reader is referred to \cite{turrigiano2008self}.

\section{Structural plasticity as growth dynamics}
\label{sec:growth_adapt}
Structural plasticity involves multiple morphological changes that happen in response to perturbations in electrphysiological activity \cite{butz2009activity}. Morphological changes include dendritic/axonal length variations (during development), synapse formation and elimination (dendritic spines and axonal boutons), and branching. In our model, such variations are captured 
by a single length parameter $L$, describing the average length of a dendritic arbor.

\begin{figure}[htpb]
	\centering \vspace*{3mm}
	\includegraphics[width=.84\columnwidth]{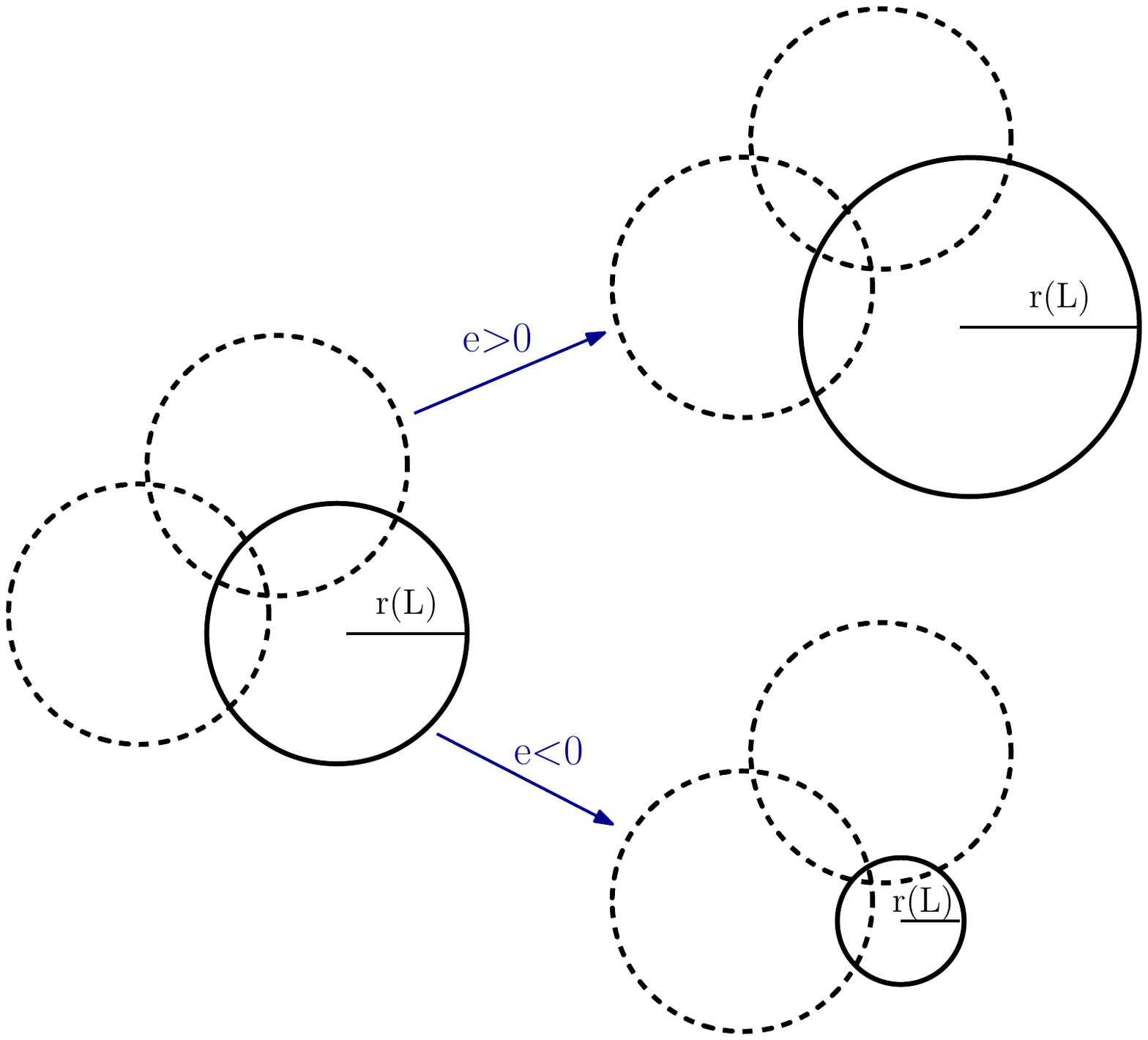}
	\caption{Guided by the mismatch $e$, as $L$ increases (decreases), the neuron's probability to form new connections increases (decreases), which is done by spatially expanding (shrinking). $r(L)$ denotes the radius of connectivity, which is a function of $L$.}
	\label{fig:conn}
\end{figure}

We model structural plasticity as a growth process that is directly coupled to the neurons's average activity,
captured by calcium concentration $[Ca^{+2}]$. The basic idea is that short dendrites have fewer connections,
thus a reduced electrical activity. Likewise, long dendrites potentially make more connections, thus enjoy stronger electrical
activity. In this sense, $L$ can also be considered as an abstract indicator of the connectivity of the neuron.
Then, a feedback mechanism adjusts $L$ to achieve homeostasis in a way that is not at all
dissimilar from synaptic scaling \cite{van2017network, kater1988calcium, tailby2005activity}:
above set-point ($e<0$), $L$ must shrink to reduce the number of synapses (pruning) /  weakening existing connections,
thus reducing the overall electrical activity; conversely, below set point ($e>0$), $L$ must increase for the neuron to 
reach out to other neurons / strengthening existing connections, ultimately increasing the level of electrical activity. This is schematically captured in Figure \ref{fig:conn}. Each neuron in the network is represented by circular neuritic field \cite{van2017network}, where the radius is changing in an activity-dependent manner (in this case the radius $r$ is a function of $L$). A connection and its strength is reflected by the size of the overlapping area. In this way, the more connections the cell receives, the stronger the average electrical activity will be, and vice versa.

We remark that our assumption about the relationship between the neuronal activity and $L$ is not about the intrinsic membrane properties but rather about the  neuron connectivity. The total excitatory input is assumed to scale with the size of the dendritic arbour. 
This is a reasonable and standard assumption \cite{van2017network}. 
One might argue that inhibitory connections develop in similar proportion, canceling out the increase in excitatory connectivity. In a mature network this may be the case, however during development, as networks are growing, excitatory connections form initially and inhibitory neurotransmitters undergo a late developmental switch \cite{ben1989giant} from being initially excitatory to inhibitory after many of the connections have formed. 

\begin{figure}[htpb]
	\centering
	\includegraphics[width=1.0\columnwidth]{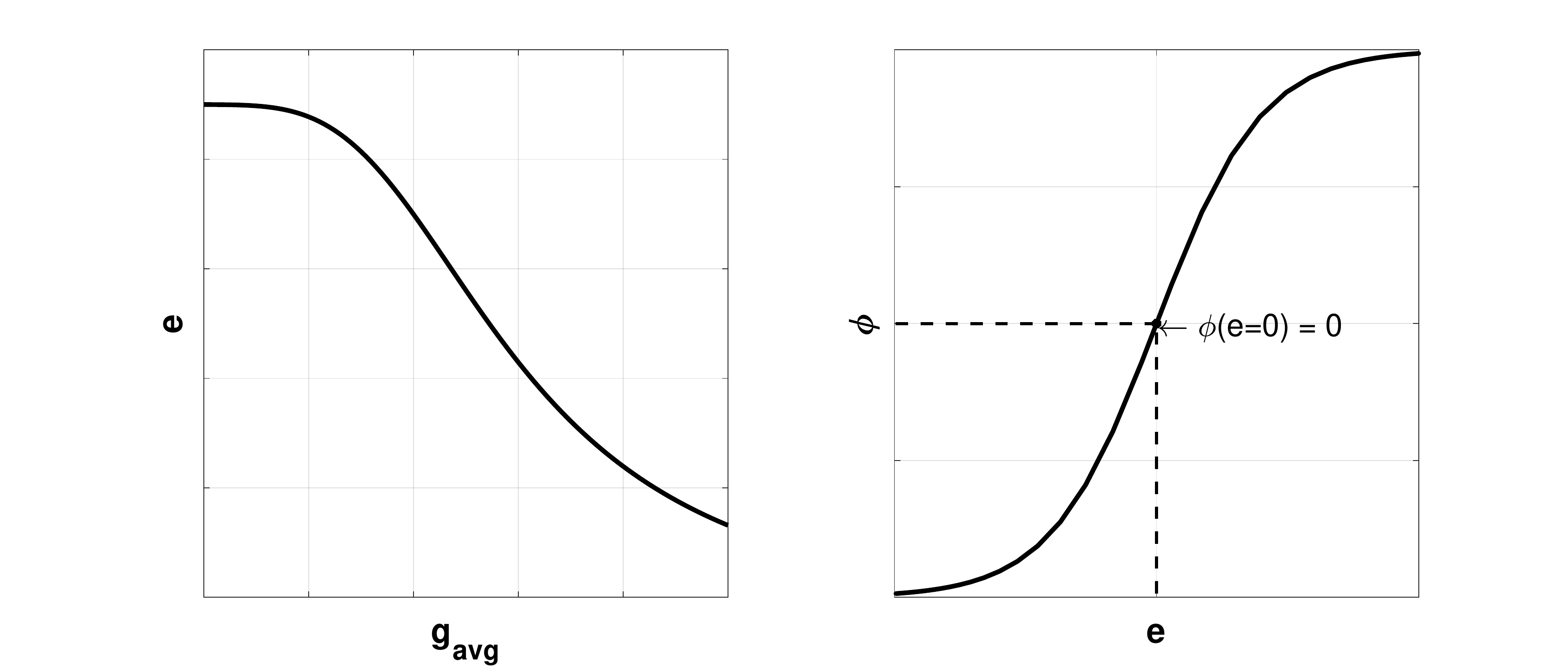}
	\caption{\textbf{Left:} shape $e = [Ca^{+2}]_{target} - h(g_{avf})$, which incorporates the static map $h$ in \eqref{EQ:ca2} (with $\beta = 5$).
		\textbf{Right:}  an example of the monotone increasing function $\phi$. The graph corresponds to \eqref{EQ:phi}.}
	\label{fig:stat_map}
\end{figure}

Based on these physiological observations, we model the growth dynamics using the nonlinear first order process
\begin{equation}
\label{EQ:l-c}
\tau \dot L  = \phi(e) -\omega_{L}L \ , \qquad
c  = \frac{L}{n} 
\end{equation}
where $\tau \gg 0$ is a slow time constant reflecting the slow dynamics of growth\footnote{
	In biological neurons, growth rates are on the order of days or weeks \cite{van1985synaptogenesis, schilling1991electrical}, while active motor-assisted transport is of the order of hours \cite{williams2016dendritic}. For example, in \textit{C. elegans}, it was found that they grow at an average rate of $0.001 \mu m/s$, while active transport rates are $\mathcal{O}(1 \mu m/s)$ \cite{brooks2017turing,sulston1977post}. 
}, and 
$\omega_{L}$ is the degradation or disassembly rate of the molecules that are responsible for synthesis of the new dendritic components, such as tubulin.
The function $\phi$, $\phi(0) = 0$, is a monotonically increasing function in the error $e$  (see Figure \ref{fig:stat_map}). 

If we interpret $L$ as a connectivity indicator,
the slope $\phi'$ captures the density of the surrounding neurons or richness of the network; a steeper $\phi$ means there are more potential connections to be made or removed, for the same amount of growth. 
We emphasize that growth in \eqref{EQ:l-c} eventually corresponds to a simple variation of capacity,
which also affects the forward and backward transport rates in \eqref{EQ:m}-\eqref{EQ:u}
 
\begin{figure}[htpb]
	\centering
	\includegraphics[width=0.97\columnwidth]{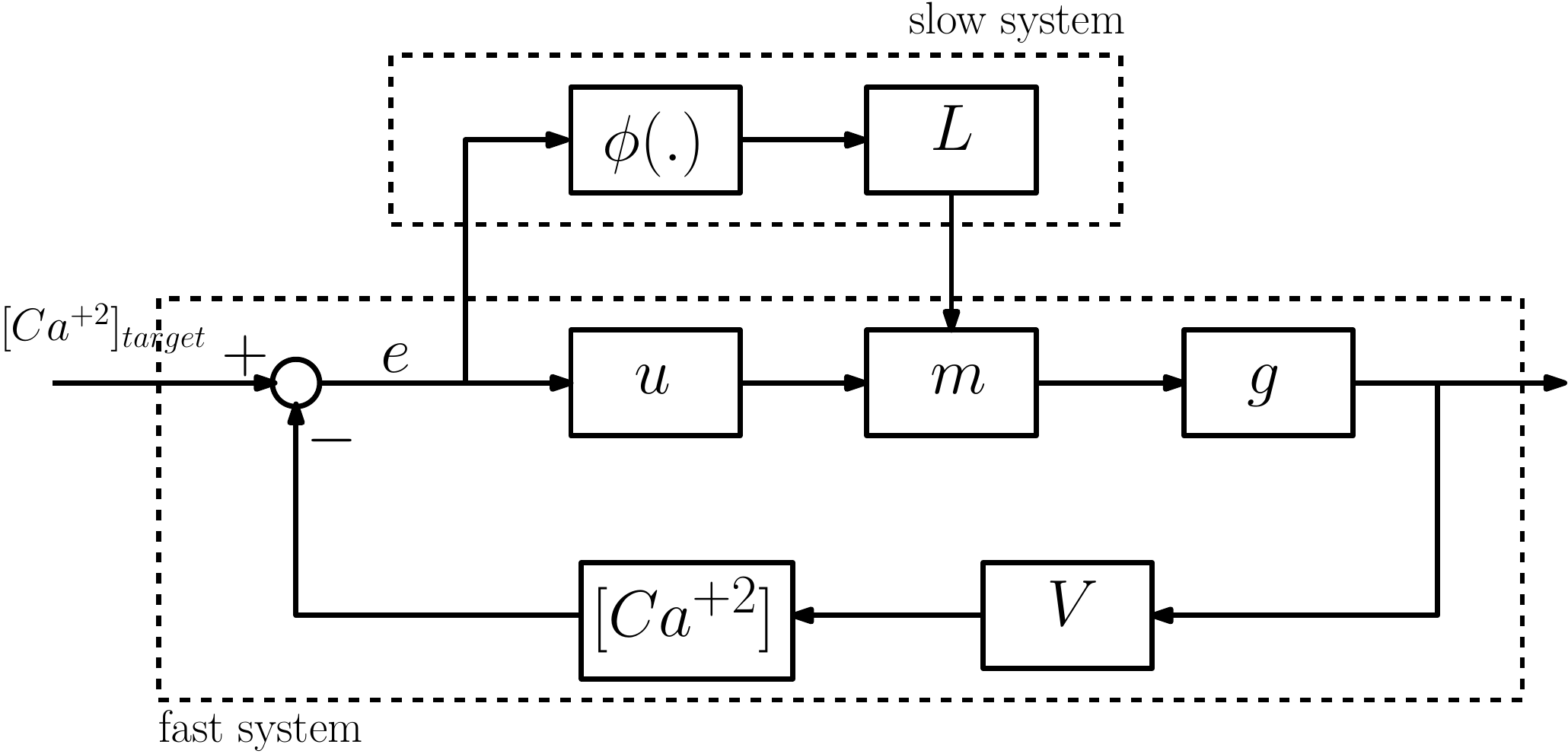}
	\caption{Complete closed loop \eqref{EQ:m}-\eqref{EQ:l-c} block diagram.}
	\label{fig:block}
\end{figure}

 Both synaptic scaling \eqref{EQ:m}-\eqref{EQ:u} and growth dynamics \eqref{EQ:l-c} aim to achieve the same objective: regulating the neuron's average activity around an (approximate) set-point. The difference is that synaptic scaling occur at a fast timescale and it works by modulating (globally) the number of ion channels $g_{i}$ in the system. In contrast, growth occurs at a slow timescale and it changes compartments' capacities to modulate the maximum allowable $g_{i}$ in each synapse. Figure \ref{fig:block} provides an illustration of the complete closed loop \eqref{EQ:m}-\eqref{EQ:l-c}.
\section{Homeostasis by fast synaptic scaling and slow growth adaptation}
\label{sec:cl}

For any fixed dendritic length $L$, synaptic scaling \eqref{EQ:m}-\eqref{EQ:u} 
is a stable process if the feedback gain $k_u$ is sufficiently small. In fact, 
the aggressiveness of the control action is fundamentally limited by the presence of transport,
typically introducing a phase lag that limits the feedback gain \cite{aljaberi2019qualitative,aljaberi2021dendritic}.
Growth adaptation is also a stable process, 
naturally occurring at a slower timescale than synaptic scaling. Thus, by timescale 
separation, the combination of synaptic scaling and growth adaptation leads
to a stable closed loop \eqref{EQ:m}-\eqref{EQ:l-c}, for a sufficiently slow growth time constant 
$\tau$.
\begin{thm}
	\label{thm}
	Under Assumption \ref{assume:feasibility},
	there exists a maximal feedback gain $\bar{k}_u>0$
	and a minimal time constant $\bar{\tau}>0$ such that,
	for every $0  \leq k_u < \bar{k}_u$, $  \tau > \bar{\tau}$, and $[Ca^{2+}]_{\text{target}}$
	the closed-loop \eqref{EQ:m}-\eqref{EQ:l-c} has a globally exponentially stable equilibrium.
\end{thm}

Taking advantage of the theoretical result in Theorem \ref{thm},
we study the system's response under different biologically relevant situations, 
to better understand the interplay between synaptic scaling and growth dynamics. 
With this aim, we set $\phi(e)$ in \eqref{EQ:l-c} as in \cite{van2017network}
\begin{align}
	\label{EQ:phi}
	\phi(e) & = 1 - \frac{2}{1 + \exp(e/\eta)} \ ,
\end{align}
and we simulate the system for the parameters  in Table \ref{table:par_val}.
\begin{table}[htbp]
	\begin{center}
		\begin{tabular}{ |c|c|c|c| } 
			\hline
			$v_f=1$ & $v_b=0.5$ & $\omega_{m} = 0. 1$& $n=2$ \\ 
			$E_{leak}=-50$ & $E_{g}=20$ &$\omega_{L} = 0.1$ &  $\eta = 0.1$ \\
			$\beta=1$ & $\alpha=1$ & $\omega_{g}=0.1$ &   $\tau = 10^{5}$  \\
			$[Ca^{2+}]_{\text{target}} = 0.5$ & $g_{leak}=0.25$ & $\omega_{u} = 10^{-5}$ &    \\
			\hline
		\end{tabular}
			\caption{} \vspace{-6mm}
				\label{table:par_val}
	\end{center}
\end{table}

The first observation is that 
\emph{growth adaptation guarantees homeostasis even if synaptic scaling is insufficient}.
First of all, note that the averaged sum of ion-channel proteins is limited by
$$g_{avg} = \frac{1}{n}\sum\limits_{i=1}^{n}g_{i}=\frac{1}{n\omega_{g}} \sum\limits_{i=1}^{n}s_{i}m_{i} \le \frac{c}{n\omega_{g}} \sum\limits_{i=1}^{n}s_{i}.$$ 
Thus, regulation is feasible 
$$0 \simeq e = [Ca^{+2}]_{target} - [Ca^{+2}] = [Ca^{+2}]_{target}-h(g_{avg})\ ,$$ 
only if the desired steady state satisfies 
\begin{equation}
\label{EQ:too_large_target}
[Ca^{+2}]_{target} \leq  h\left(\frac{c}{n\omega_{g}} \sum\limits_{i=1}^{n}s_{i}\right) \ .
\end{equation}
Inequality \eqref{EQ:too_large_target} fundamentally relates the calcium target / the desired level neural activity
to the morphological parameter $c = L/n$. It shows that, without growth adaptation, high levels of neural activity
($[Ca^{+2}]_{target}$ large) cannot be attained in closed loop because of the finite capacity of cargo transport. However, taking advantage of growth adaptation, the neuron can develop its morphology to reach the desired calcium target. 

\begin{figure}[htbp]
	\begin{center}
		\includegraphics[width=1.0\columnwidth]{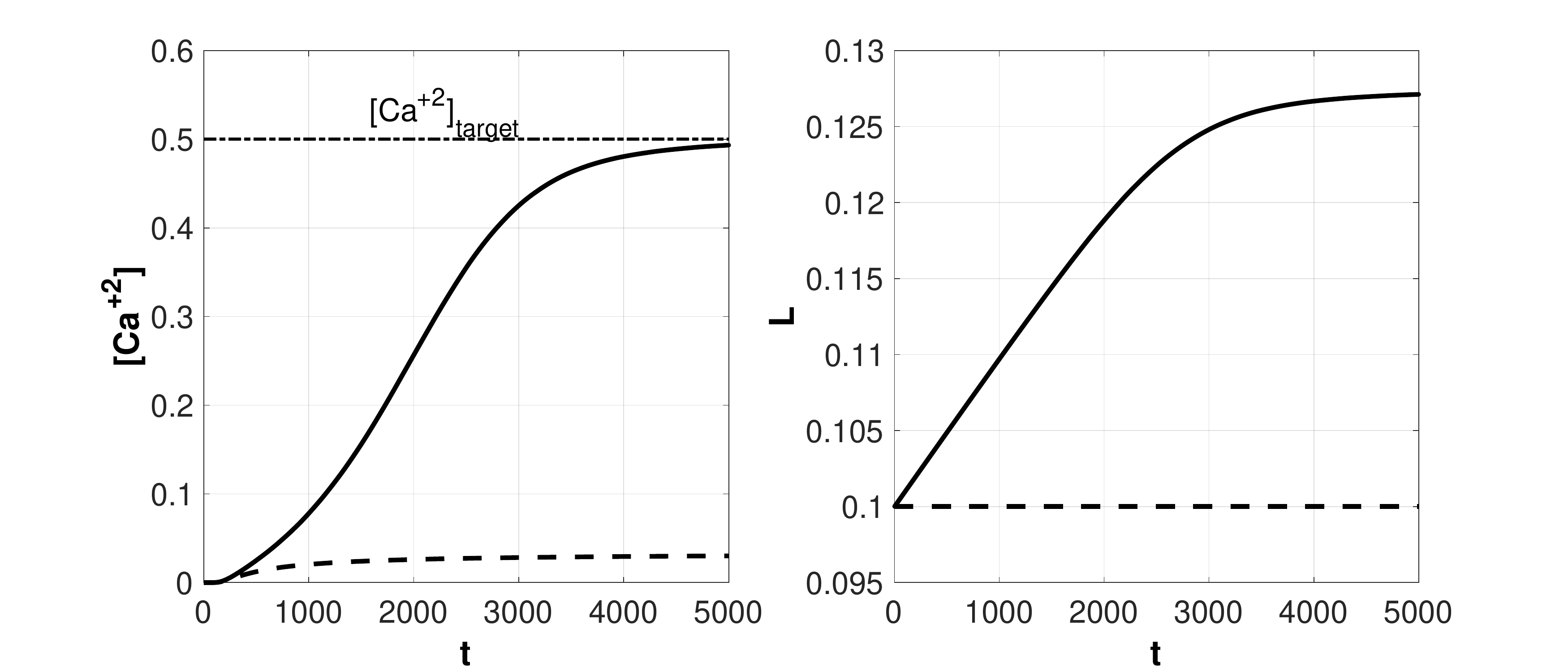}
		\caption{Average activity and length for the synaptic scaling model \eqref{EQ:m}-\eqref{EQ:u} (dashed) and synaptic scaling model with growth dynamics \eqref{EQ:m}-\eqref{EQ:l-c} (solid). $k_{u} = 0.001$ and $L_{0} = 0.1$.}
		\label{fig:R1}
	\end{center}
\end{figure}

These two cases are illustrated through simulation, within a comparison between 
synaptic scaling without growth adaptation  \eqref{EQ:m}-\eqref{EQ:u},
and synaptic scaling with growth adaptation  \eqref{EQ:m}-\eqref{EQ:l-c}.
Results are summarized in Figure \ref{fig:R1}. Left and right graphs shows
the calcium $[Ca^{+2}]$ trajectory and the length $L$ trajectory, respectively.
Dashed lines correspond to synaptic scaling without adaptation,
while continuous lines correspond to the growth adaptation case. 
Figure \ref{fig:R1} shows the case in which $[Ca^{+2}]_{target} = 0.5$ is not compatible with 
the the initial capacity $c = L/n = 0.1/n$ in the sense of \eqref{EQ:too_large_target}. 
The dashed line shows that synaptic scaling without growth adaptation is stable but far
from target. This is not the case of synaptic scaling with growth adaptation,
whose calcium trajectory asymptotically converges to $[Ca^{+2}]_{target}$,
taking advantage of the increased average length $L$, thus of larger capacity $c$.

Considering $L$ as a connectivity parameter, 
the biological interpretation is that 
the neuron is below its target activity level and therefore attempts to increase its activity  by extending its dendritic tree to form new connections. Likewise, considering $L$ as a morphological parameter, 
the neuron increases the size of its spines to allow more ion channels to flow to the synapse, which also increases the electrical activity. 

\begin{figure}[htbp]
	\begin{center}
		\subfigure[Synaptic scaling model \eqref{EQ:m}-\eqref{EQ:u}]
		{
			\includegraphics[width=1.0\columnwidth]{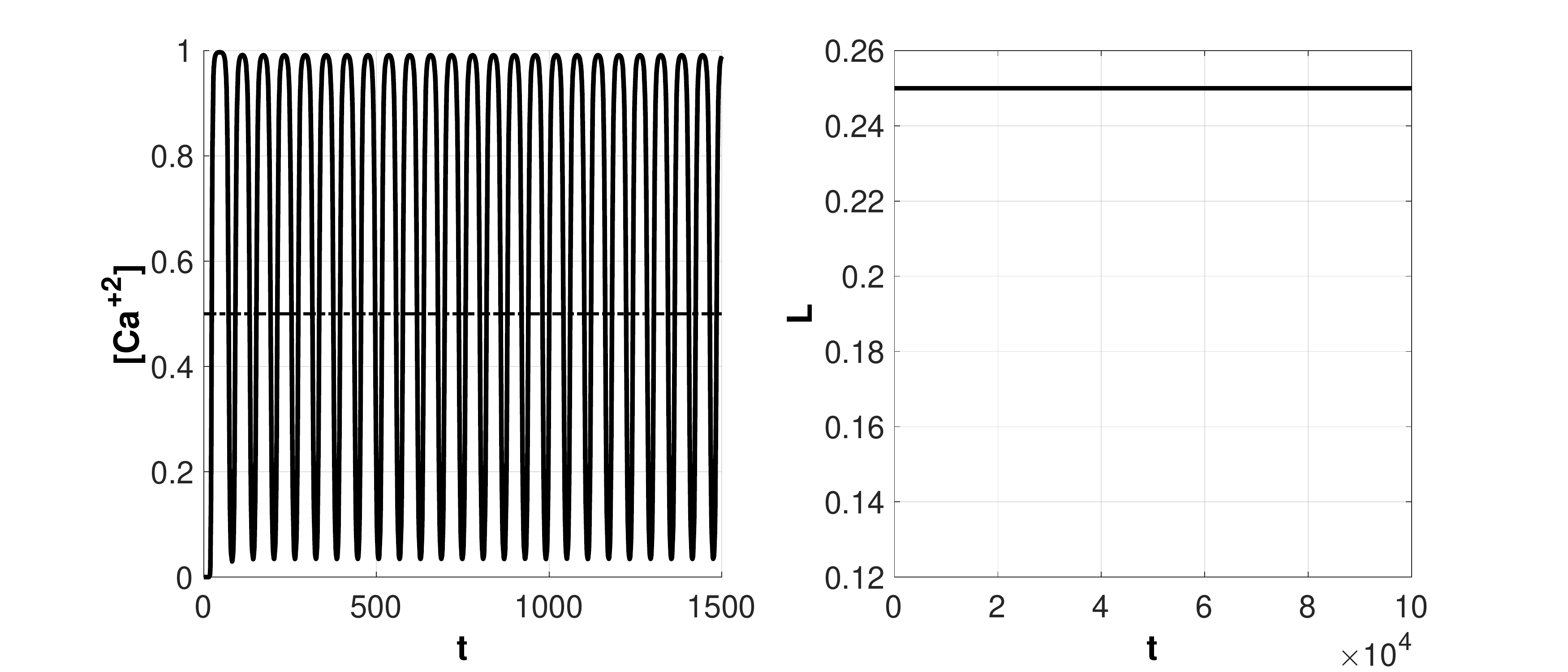}
			\label{fig:R2-a} 
		}
		\subfigure[Synaptic scaling model with growth dynamics \eqref{EQ:m}-\eqref{EQ:l-c}]
		{
			\includegraphics[width=1.0\columnwidth]{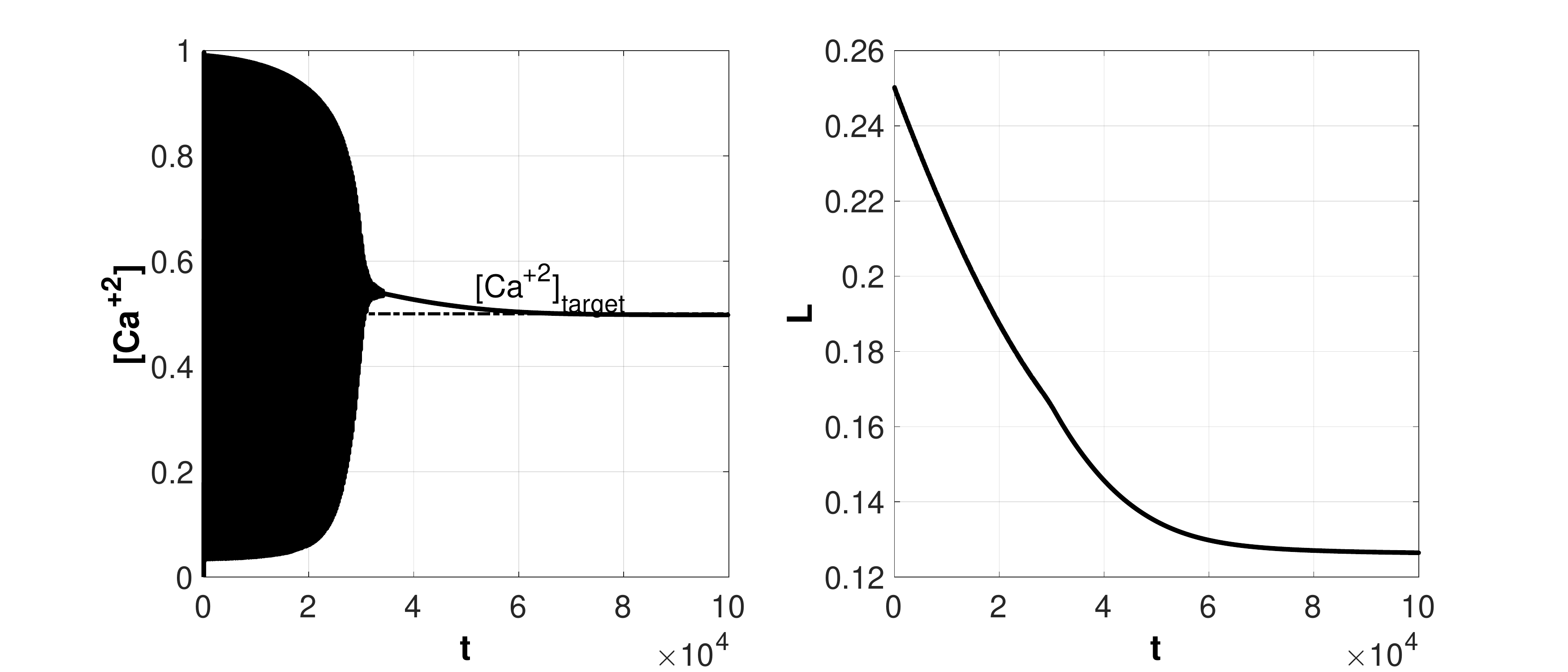}
			\label{fig:R2-b}
		} 		\vspace{-2mm}
		\caption{ grow adaptation \eqref{EQ:l-c} increases the maximum allowable $\bar{k}_{u}$. 		
			Simulations were done with $k_{u} = 0.01$ and $L_{0} = 0.25$. For readability, the calcium trajectory in 
			the left graph of Figure \ref{fig:R2-a} is represented on the reduced domain $0\leq t \leq 1500$.}
		\label{fig:R2}
	\end{center}
\end{figure}

The second observation, derived from simulations, is that \emph{growth adaptation may compensate for pathological oscillations, enabling more aggressive synaptic scaling}. 
Aggressive feedback gains $k_u$ may lead to pathological oscillations in synaptic scaling \cite{aljaberi2019qualitative,aljaberi2021dendritic}, as shown in Figure  \ref{fig:R2-a}. However,
these oscillations are dampened through growth adaptation, as shown in Figure  \ref{fig:R2-b}, 
reaching the desired set-point. 
The intuition is that  \eqref{EQ:l-c} is essentially a low pass filter therefore it filters calcium oscillations, extracting the oscillations bias. The overall growth adaptation is thus driven by this bias. When the bias is above the desired calcium target, as in Figure  \ref{fig:R2},  the average length will reduce, stabilizing the oscillations.
The biological interpretation is that large neurons reduces their size when their average electrical activity is irregular (oscillatory).

\begin{figure}
	\begin{center}
		\subfigure[$\tau = 10^{4}$]
		{
			\includegraphics[width=1.0\columnwidth]{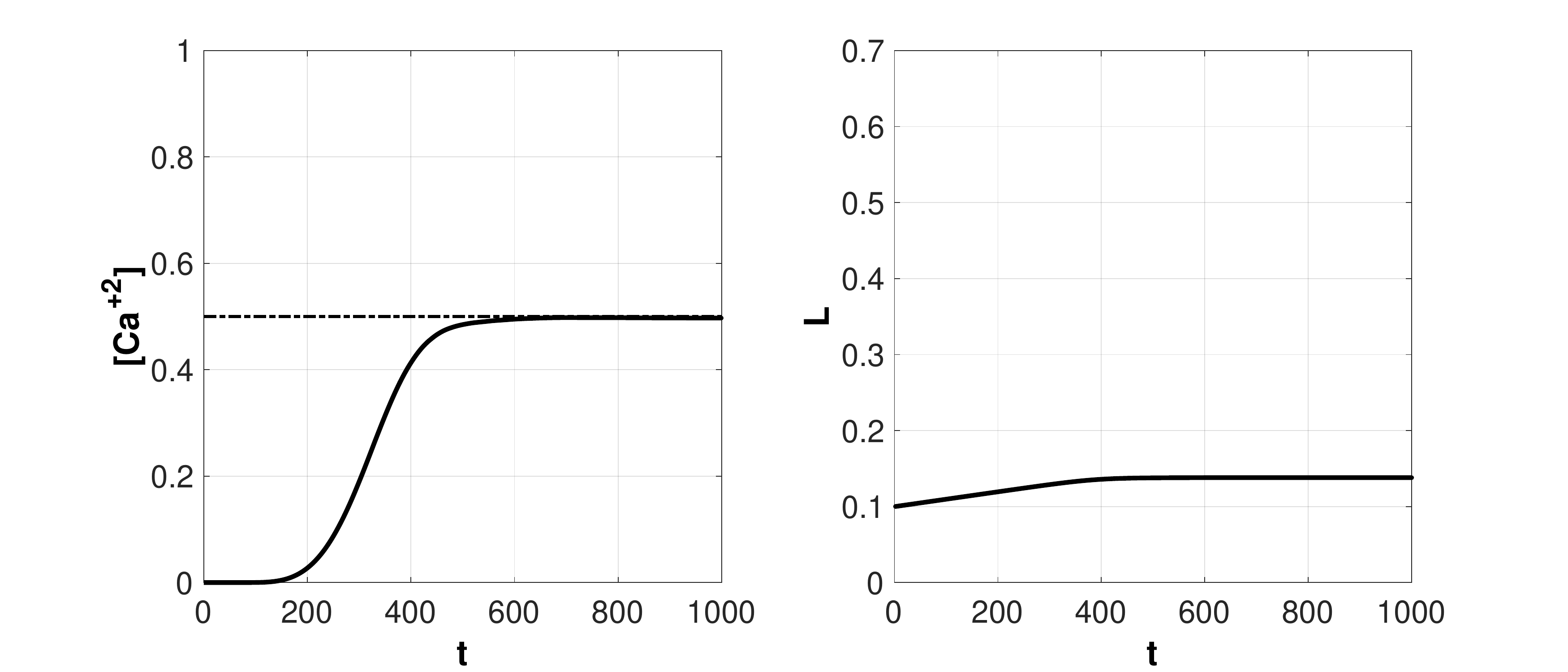}
			\label{fig:R3-a}
		}
		\subfigure[$\tau = 10^{3}$.]
		{
			\includegraphics[width=1.0\columnwidth]{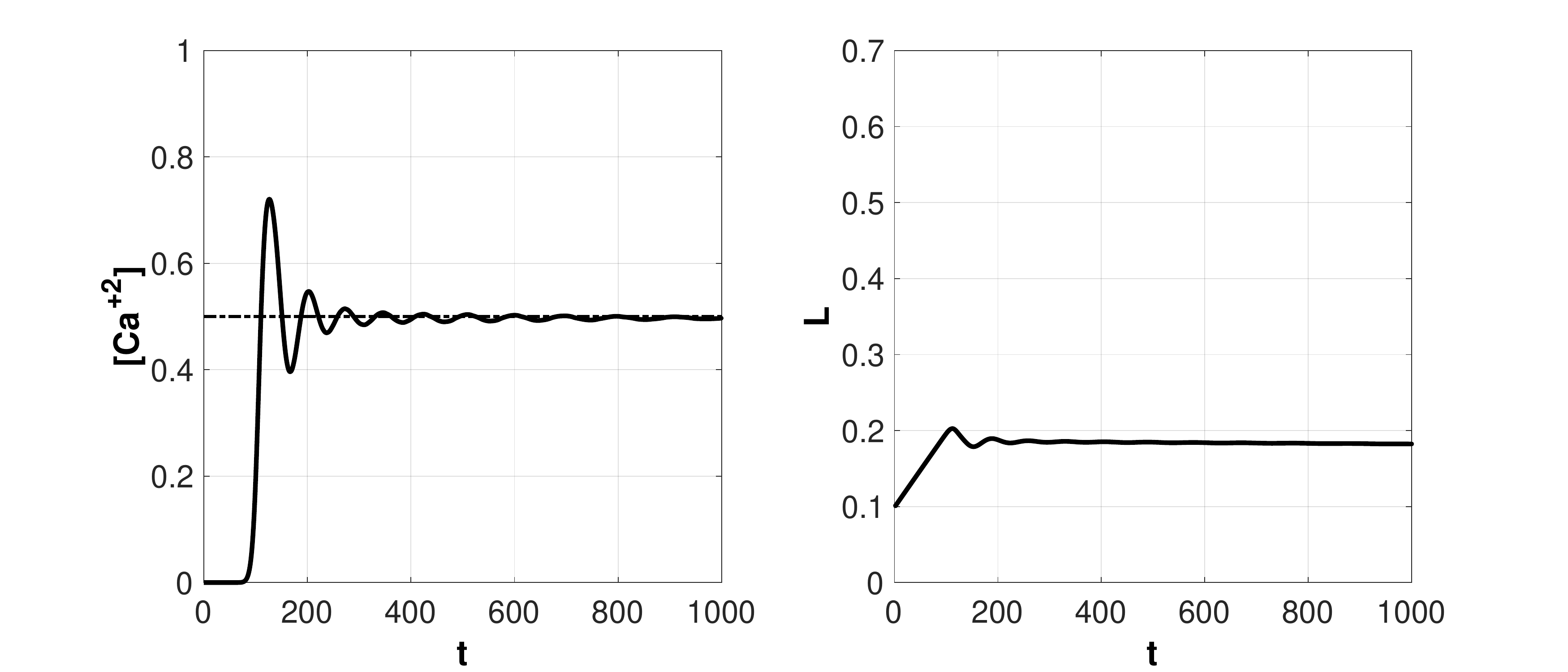}
			\label{fig:R3-b}
		}
		\subfigure[$\tau = 10^{2}$.]
		{
			\includegraphics[width=1.0\columnwidth]{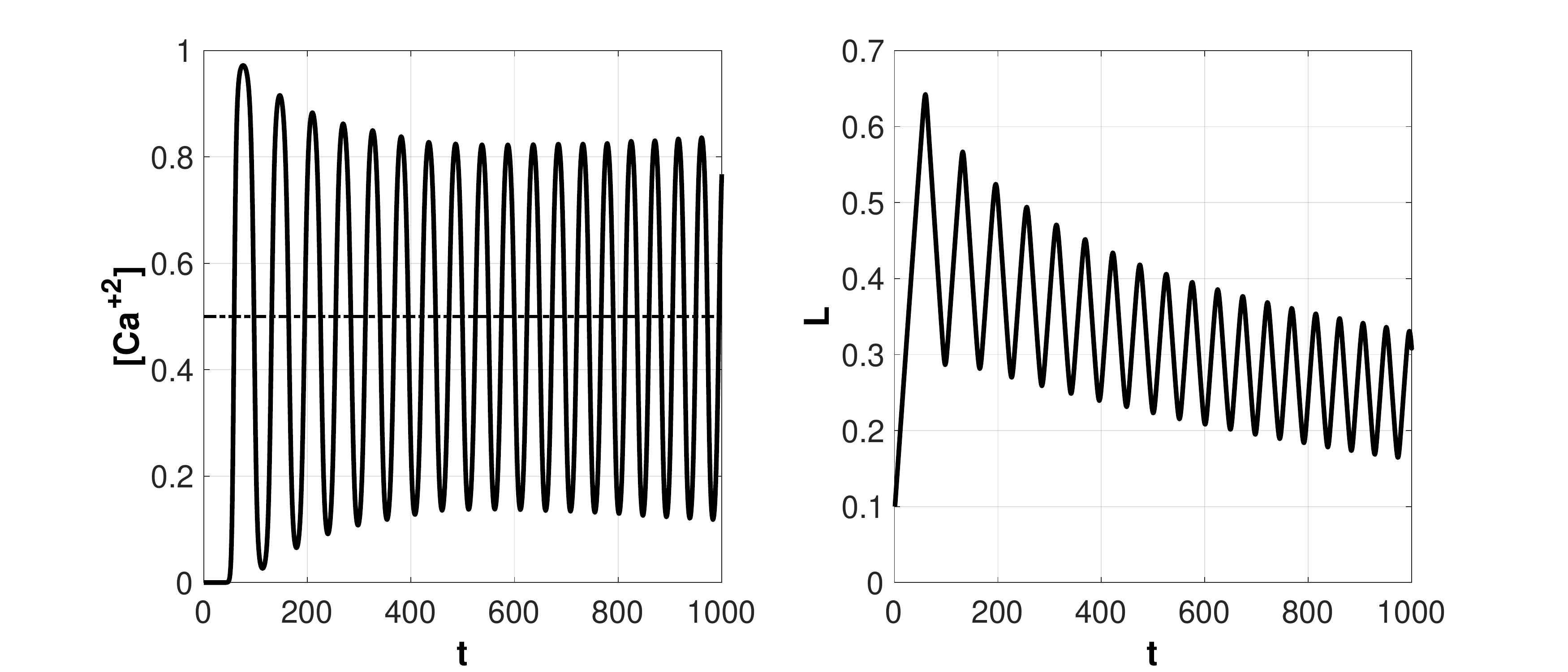}
			\label{fig:R3-c}
		}\vspace{-2mm}
		\caption{Inadequate timescale separation introduces fragility.}
		\label{fig:R3}	
	\end{center}
\end{figure}

The last observation is that \emph{inadequate timescale separation leads to fragility}.
Theorem \ref{thm} guarantees closed loop stability under the strong hypothesis of 
timescale separation between synaptic scaling and growth adaptation.
The simulations in Figure \ref{fig:R3} shows that timescale separation is actually needed
for stability. As $\tau$ decreases the system stability becomes more fragile.
Reducing $\tau$ produces damped oscillations and a further reduction
eventually leads to sustained oscillations, for smaller values of $\tau$. 
\section{conclusions} 
We presented a model of nonlinear dendritic trafficking with growth adaptation to 
study two distinct  homeostatic mechanisms: synaptic scaling and structural plasticity. We studied the interplay between the two and how timescale separation provides the means to improve the overall performance of the system. Using contraction arguments combined with singular perturbation theory, we proved exponential stability 
of the closed-loop equilibrium and we discussed several features of the closed-loop system, supported by 
simulations.
Our growth model is very simple. Future research will focus on extended modeling of growth.

\bibliographystyle{IEEEtran}
\bibliography{ECC_2021_ref}

\section*{Appendix: Proof of theorem \ref{thm}}
We prove the stability of the closed loop system \eqref{EQ:m}-\eqref{EQ:l-c}
by singular perturbation arguments, specifically invoking \cite[Theorem 11.4]{khalil2002nonlinear}.
To use this theorem, we prove the exponential stability of the equilibrium of the so-called boundary layer, or fast system. This is constructed from  \eqref{EQ:m}-\eqref{EQ:u}.
Furthermore we prove the exponential stability of the equilibrium of the so-called reduced system,
constructed from \eqref{EQ:l-c}, by relaxing the fast system at steady state. 
Part 1 and Part 2 below show that the fast system and the reduced system are both exponentially contractive systems, which entail exponential stability of their respective equilibria.
In particular, the fast system is contractive for any feedback gain $0 \leq k_u < \bar{k}_u$, provided that
$\bar{k}_u$ is sufficiently small. 
Thus, stability of the closed loop \eqref{EQ:m}-\eqref{EQ:l-c} follows from \cite[Theorem 11.4]{khalil2002nonlinear}, under the assumption of sufficient time-scale separation $\tau \gg1$.

\noindent\emph{Part 1: contraction / stability of the fast system}

 First we multiply equations \eqref{EQ:m}-\eqref{EQ:l-c} by $\epsilon = \frac{1}{\tau}$. We start by proving the stability of the fast system \eqref{EQ:m}-\eqref{EQ:u}. The time derivative in the equations below refers to the scaled time $ \tilde{t} = \frac{t}{\tau}$. In the fast timescale, the slow variable $L$ is considered as constant.
The linearized dynamics of the time-scaled fast system \eqref{EQ:m}-\eqref{EQ:u} reads
\begin{align}
\label{EQ:fast_partial}
\epsilon \delta \dot  m & = \frac{\partial f}{\partial m}\delta m  +B \delta u  \\
\epsilon \delta \dot  g & = S\delta m - \Omega_{g} \delta g \nonumber\\
\epsilon \delta \dot  u & = -k_{u}\frac{\partial h }{\partial g}\left(\frac{\mathbf{1}^{T}g}{n}\right)\frac{\mathbf{1}^{T}}{n}\delta g - \omega_{u} \delta u. \nonumber 
\end{align}
where $f(m,L)$ is the right-hand side of \eqref{EQ:m}, $S=\text{diag}\{s_{i}\}$, $\Omega_{g}=\text{diag}\{\omega_{g}\}$, $m=[m_{0},\dots,m_{n}]^{T}$, $g=[g_{1},\dots, g_{n}]^{T}$ and 
$g_{avg} = \frac{1}{n} \sum_i^n g_i = \frac{\mathbf{1}^T g}{n}$, where $\mathbf{1}$ is
a vector of ones.

We need to show that \eqref{EQ:fast_partial} is  a contracting system, which implies the existence
of a globally exponentially stable equilibrium when the contracting distance is a norm.
We first note that $ \frac{\partial f}{\partial m}^{T} + \frac{\partial f}{\partial m}  \le - 2\omega_{m} I < 0$. Take the differential Lyapunov function $V = \frac{\rho_{m}}{2}\delta m ^{T} \delta m + \frac{\rho_{g}}{2}\delta g ^{T} \delta g + \frac{1}{2}\delta u ^{T} \delta u$. The coefficients
$\rho_m > 0 $ and $\rho_g > 0 $ will be defined later.
Its time derivative reads
\begin{align}
\label{EQ:V_dot}
\dot  V & = \dot  V_{m} +  \dot  V_{g}  + \dot  V_{u}
\end{align}
where 
\begin{align}
\dot  V_{m}  & = \frac{\rho_{m}}{2}\!\left(\left[\frac{\partial f}{\partial m}\delta m + B \delta u  \right]^{T}\!\!\!\!\delta m + \delta m^{T} \!\!\left[\frac{\partial f}{\partial m}\delta m + B \delta u  \right]\right)  \nonumber \\
& < -\rho_{m}\omega_{m}\delta m^{T}\delta m +\rho_{m}B\delta u^{T} \delta m \ , \nonumber
\end{align}
\begin{align}
\dot  V_{g} 
&= \rho_{g}S\delta m^{T}\delta g - \rho_{g}\Omega_{g}\delta g^{T}\delta g\nonumber, 
\end{align}
and
\begin{align}
\dot  V_{u} 
&=  -k_{u}\frac{\partial h }{\partial g}\left(\frac{\mathbf{1}^{T}g}{n}\right)\frac{\mathbf{1}^{T}}{n}\delta g\delta u - \omega_{u} \delta u^{T}\delta u. \nonumber
\end{align}
Therefore, \eqref{EQ:V_dot} satisfies 
\begin{align}
\label{EQ:V_dot_LMI}
\dot V & < -\rho_{m}\omega_{m}\delta m^{T}\delta m + \rho_{m}B\delta u^{T} \delta m + \rho_{g}S\delta m^{T}\delta g  \\
& - \rho_{g}\Omega_{g}\delta g^{T}\delta g -k_{u}\frac{\partial h }{\partial g}\left(\frac{\mathbf{1}^{T}g}{n}\right)\frac{\mathbf{1}^{T}}{n}\delta g\delta u - \omega_{u} \delta u^{T}\delta u \nonumber\\
& < -\rho_{m}|\omega_{m}||\delta m|^{2} + \rho_{m} |B||\delta u||\delta m| + \rho_{g}|S||\delta m||\delta g|\nonumber \\
& -\!\rho_{g} \lambda_{min}(\Omega_{g})|\delta g|^{2}\! +\!k_{u}\left|\frac{\partial h }{\partial g}\!\left(\!\frac{\mathbf{1}^{T}\!g}{n}\!\right)\!\frac{\mathbf{1}^{T}}{n}\right|\!|\delta g||\delta u|\! -\! \omega_{u} |\delta u|^{2}\!. \nonumber
\end{align}
The right-hand side of \eqref{EQ:V_dot_LMI} is bounded by
\resizebox{1\columnwidth}{!}{
	\begin{minipage}{\columnwidth}
		\begin{align*}
		\begin{bmatrix}
		|\delta m| \\
		|\delta g| \\
		|\delta u|
		\end{bmatrix}^{T}
		\underbrace{\begin{bmatrix}
			-\rho_{m}|\omega_{m}| & \frac{1}{2}\rho_{g}|S| &\frac{1}{2} \rho_{m} |B|\\
			\frac{1}{2}\rho_{g}|S| & -\rho_{g} \lambda_{min}(\Omega_{g}) & \frac{1}{2}k_{u}\left|\frac{\partial h }{\partial g}\left(\frac{\mathbf{1}^{T}g}{n}\right)\frac{\mathbf{1}^{T}}{n}\right| \\
			\frac{1}{2} \rho_{m} |B| & \frac{1}{2}k_{u}\left|\frac{\partial h }{\partial g}\left(\frac{\mathbf{1}^{T}g}{n}\right)\frac{\mathbf{1}^{T}}{n}\right| & - \omega_{u}
			\end{bmatrix}}_{-Q}
		\begin{bmatrix}
		|\delta m| \\
		|\delta g| \\
		|\delta u|
		\end{bmatrix}.
		\end{align*}
	\end{minipage}
}
Next we show that $Q > 0$, using the Sylvester criterion.
This guarantees contraction, therefore global exponential stability of the fast system equilibrium
We start by finding conditions under which the leading principal minors of 
\begin{align}
\begin{bmatrix}
\rho_{m}|\omega_{m}| & -\frac{1}{2}\rho_{g}|S| &-\frac{1}{2} \rho_{m} |B|\\
-\frac{1}{2}\rho_{g}|S| & \rho_{g} \lambda_{min}(\Omega_{g}) & -\frac{1}{2}k_{u}\left|\frac{\partial h }{\partial g}\!\left(\frac{\mathbf{1}^{T}g}{n}\right)\!\!\frac{\mathbf{1}^{T}}{n}\right| \\
-\frac{1}{2} \rho_{m} |B| & -\frac{1}{2}k_{u}\left|\frac{\partial h }{\partial g}\!\left(\frac{\mathbf{1}^{T}g}{n}\right)\!\frac{\mathbf{1}^{T}}{n}\right| &  \omega_{u} 
\end{bmatrix}
\end{align}
are positive. We will use the following facts:
 $|\omega_{m}|=\omega_{m}$, $|S| = s$, where $s =  \max\limits_{i} \{ s_{i} \}$, $\lambda_{min}(\Omega_{g}) = \omega_{g}$, $|B| = 1$.

The first principal minor must satisfy $ \rho_{m}\omega_{m} > 0$, which is true.
The second principal minor must satisfy
\begin{align}
\label{EQ:cond1}
\rho_{m}\rho_{g} \omega_{m}\omega_{g} - \frac{1}{4}\rho^{2}_{g}s^{2} 	> 0.
\end{align}
The last principal minor must satisfy
\begin{align*}
\rho_{m}\rho_{g}\omega_{g}\omega_{g}\omega_{u} -\frac{1}{4}\rho_{g}\rho_{m}sk_{u}\left|\frac{\partial h }{\partial g}\left(\frac{\mathbf{1}^{T}g}{n}\right)\frac{\mathbf{1}^{T}}{n}\right| + &
\\
-\frac{1}{4}\rho^{2}_{m}\rho_{g}\omega_{g} -\frac{1}{4}\rho_{m}\omega_{m}k^{2}_{u}\left|\frac{\partial h }{\partial g}\!\left(\!\frac{\mathbf{1}^{T}g}{n}\!\right)\!\frac{\mathbf{1}^{T}}{n}\right|^{2} \!\! -\! \frac{1}{4}\rho^{2}_{g}s^{2}\omega_{u} & > 0 
\end{align*}
which can be re-arranged as
\begin{align}
& \rho_{m}\rho_{g}\omega_{g}\omega_{g}\omega_{u} - \frac{1}{4}\rho^{2}_{g}s^{2}\omega_{u}  
 - \frac{1}{4}\rho^{2}_{m}\rho_{g}\omega_{g}  \nonumber \\
> \ & \frac{k_u\rho_m}{4} \left|\frac{\partial h }{\partial g}\!\left(\frac{\mathbf{1}^{T}g}{n}\right)\!\frac{\mathbf{1}^{T}}{n}\right| \left( \rho_{g}s 
+ \omega_{m}k_{u}\left|\frac{\partial h }{\partial g}\!\left(\frac{\mathbf{1}^{T}g}{n}\right)\!\frac{\mathbf{1}^{T}}{n}\right|\right) 
\label{EQ:cond2}
\end{align}
In order for the above inequality to hold, we need the left-hand side 
to be positive and larger than the right hand side. So, 
\eqref{EQ:cond1} and  \eqref{EQ:cond2} hold if we select 
\begin{enumerate}
	\item $\rho_{g} < \frac{2\rho_{m} \omega_{m}\omega_{g}}{s^2}$.
	\item $\rho_{m} < 2\omega_{u}\omega_{m}$.
	\item $0 \leq k_{u} < \bar{k}_u$ is sufficiently small.
\end{enumerate}
Under these conditions, 
$\dot{V} \leq -\bar{\lambda} V$ for some $\bar{\lambda}>0$.

The exponential decay of the differential Lyapunov function guarantees global incremental exponential stability of the fast system, \cite[Theorem 1]{Forni2013differential}.
This implies global exponential stability of the equilibrium of the fast system.

\vspace{5mm}
\noindent\emph{Part 2: contraction / stability of the reduced system}

We study the stability of the reduced system given by \eqref{EQ:l-c}
for $e$ computed from the fast system at steady state. 
Thus, as a first step, we study the monotonicity properties of the static relationship between $e$ and $L$,
denoted by $e = r(L)$. 

Define $M := \sum\limits_{i=1}^{n}m_{i}$ and $G := \sum\limits_{i=1}^{n}g_{i}$.
At steady state, $ \dot{m}_0 = 0$, $\dot{M} =0$,  $\dot{G} = 0$,  $\dot{u}  = 0$,
we have 
\begin{equation}
\left\{\begin{split}
0  & = u - m_{0} \left(\frac{L}{n} - pM\right)-\omega_{m}m_{0}  \\
0 & =  m_{0}\left(\frac{L}{n} - pM\right) -\omega_{m}M  \\
0 & = sM - \omega_{g}G \\
0 & = k_{u}e - \omega_{u} u  \ , 
\end{split}\right.
\label{EQ:M_dyn}
\end{equation}
where we have written $m_1$ at steady state as $m_1 = p M$ with $0<p<1$. 
For simplicity, we use $x$ to denote the vector $x =[m_{0}; M; G; u]^{T}$, and $R(x,L)$ to 
denote the right-hand side of \eqref{EQ:M_dyn}.

The monotonicity of the static relationship $e = r(L)$ can be determined
from the equation $R(x,L) = 0$. For instance, 
$
\frac{ \partial R}{\partial x}\delta x +  \frac{ \partial R}{\partial L}\delta L  = 0 \,, 
$
which gives
\begin{equation}
\label{EQ:SM}
\delta x  = - \left[\frac{ \partial R}{\partial x}\right]^{-1}\frac{ \partial R}{\partial L}\delta L \ .
\end{equation}
We observe that the inverse $\left[\frac{ \partial R}{\partial x}\right]^{-1}$ 
must exists since the fast system is contractive.
Furthermore, 
the error $e = [Ca^{+2}]_{target} - h(G/n) =: E(x)$.
Thus, we get
\begin{equation}
\label{EQ:SM_e}
\delta e 
=  \frac{\partial E}{\partial x} \delta x 
= \underbrace{- \frac{\partial E}{\partial x} \left[\frac{ \partial R}{\partial x}\right]^{-1}\frac{ \partial R}{\partial L}}_{\partial r/ \partial L }\delta L \ .
\end{equation} 

We observe that 
$\frac{\partial E}{\partial x} = [0 \,  0 \, \frac{\partial E}{\partial x_{3}} \, 0]$
and that $\frac{\partial E}{\partial x_{3}} < 0$ (as shown in Figure \ref{fig:stat_map}-left). 
Computing explicitly \eqref{EQ:SM_e} we get 
\begin{equation}
\label{EQ:de_dL}
\frac{\partial r}{\partial L}  = \frac{\mu_{1}}{\mu_{2}\mu_{4} + \mu_{3}} \ ,
\end{equation}
where
\begin{equation}
\begin{split}
\mu_{1} & = s\omega_{m}\omega_{u}m_{0}\frac{\partial E}{\partial x_3} < 0 \\\
\mu_{2} & = \omega_{m}\omega_{g}\omega_{u} - k_{u}s\frac{\partial E}{\partial x_3} > 0  \\
\mu_{3} & = n\omega_{m}\omega_{g}\omega_{u} (\omega_{m}+ m_{0}p) > 0  \\
\mu_{4} & = L - npM > 0 \ .
\end{split}
\label{EQ:de_dL_coeff}
\end{equation}
The latter inequality follows from
$$
npM = \frac{L}{c}\times\frac{m_{1}}{M}\times M = \frac{m_{1}}{c}L < L \ .
$$
Thus, from  \eqref{EQ:de_dL} and \eqref{EQ:de_dL_coeff},
we get $\frac{\partial r}{\partial L} < 0 $ for any $L > 0$. 

From the argument above we conclude that $e=r(L)$ is strictly
decreasing. This feature can be used to show contraction of the
reduced system. The reduced system and its linearization read
\begin{equation}
\begin{split}
\label{EQ:BL}
\dot L & =   \phi(r(L)) -\omega_{L}L 	\\
\delta \dot  L & = \underbrace{\frac{\partial \phi }{\partial e}}_{>0} \underbrace{\frac{\partial r }{\partial L}}_{<0}\delta L - \omega_{L}\delta L \ . 
\end{split}
\end{equation}
\eqref{EQ:BL} is a contractive dynamics, thus 
the equilibrium of the reduced system is globally exponentially stable.
\end{document}